\documentclass[12pt]{article}
\usepackage{}
\usepackage{geometry}             
\usepackage{graphicx}
\usepackage{amssymb}
\usepackage{amsmath}
\usepackage{epstopdf}
\usepackage{comment}
\usepackage{cite}
\usepackage{caption}
\usepackage{arcs}

\usepackage[hyperindex=true,
          pdfstartview=FitH,
          bookmarksnumbered=true,
          bookmarksopen=true,
          citecolor=blue,
          linkcolor=blue,
          colorlinks=true,
          pdfborder=001,
          unicode]{hyperref}

\allowdisplaybreaks

\newcommand{\rd}{{\rm d}}

\newcommand{\pfrac}[2]{\left(\frac{\partial #1}{\partial #2}\right)}
\newcommand{\bfrac}[2]{\left(\frac{#1}{#2}\right)}
\newcommand{\nbfrac}[2]{\left(-\frac{#1}{#2}\right)}

\newcommand{\blue}[1]{\textcolor{blue}{#1}}
\renewcommand{\blue}[1]{\textcolor{black}{#1}}  

\begin{document}
\title{Restricted phase space thermodynamics of charged
AdS black holes in conformal gravity}
\author{Xiangqing Kong\thanks{
{\em email}: \href{mailto:2120200165@mail.nankai.edu.cn}{2120200165@mail.nankai.edu.cn}},~
Zhiqiang Zhang\thanks{{\em email}: \href{mailto:2120210176@mail.nankai.edu.cn}
{2120210176@mail.nankai.edu.cn}}~
and Liu Zhao\thanks{Correspondence author, {\em email}: 
\href{mailto:lzhao@nankai.edu.cn}{lzhao@nankai.edu.cn}}\\
School of Physics, Nankai University, Tianjin 300071, China}

\date{}
\maketitle
\begin{abstract}
The thermodynamics of charged spherically symmetric 
AdS black holes in conformal gravity is 
revisited using the recently proposed restricted phase space (RPS) formalism. 
This formalism avoids all the bizarreness that arose in the extended phase space 
formalism for this model. It is found that the charged AdS black holes in this 
model may belong to a new universality class which is different from all 
previously studied cases under the RPS formalism. 
Besides the distinguished isocharge $T-S$ and isothermal $\Phi-Q_e$ behaviors, 
the absence of Hawking-Page transition is another notable feature. 
On the other hand, in the high temperature limit, the thermodynamic behavior of the 
present model become exactly the same as that of the Einstein gravity and the 
black hole scan models, which add further evidence for the universality of the 
recently reported correspondence between high temperature AdS black holes and 
low temperature quantum phonon gases in nonmetallic crystals.
\end{abstract}

\newpage

\section{Introduction}

After more than one hundred years from the date of its birth, Einstein gravity 
remains to be the most promising theory of relativistic gravity. 
However, there are several reasons to look at alternative theories of gravity, 
for instance, the need for a renormalizable quantum theory of gravity, the interpretation
of the expanding universe and the galaxy rotation curves, etc. 
Among the numerous choices of extended theories of gravity, conformal gravity
has attracted considerable interests, partly because of its on-shell equivalence to 
Einstein gravity and its power counting renormalizability \cite{Maldacena:2011mk}, 
and partly because of the fact that its spherically symmetric solution 
contains a term linear in radial coordinate which may play some role in 
explaining the galaxy rotation curves
as an alternative for dark matter \cite{Mannheim,nesbet}. 
As a physically viable model of relativistic gravitation, 
it is natural to look into more detailed behaviors of conformal gravity, 
and the study of its black hole solutions and 
the corresponding thermodynamic properties \cite{Lu:2011zk,Li:2012gh,Lu:2012xu} 
is of great importance in this regard.

There are several formalisms for studying black hole thermodynamics. 
The traditional formalism focuses on the initial establishment of the basic 
laws of black hole thermodynamics and the calculation of thermodynamic quantities
\cite{bekenstein1972black, bekenstein1973black, bardeen1973four, 
hawking1975particle, bekenstein1975statistical}. A relatively modern formalism 
known as the extended phase space (EPS) formalism appeared about fifteen years ago
\cite{kastor2009enthalpy}, which takes the negative cosmological constant 
(hence only applicable to AdS black holes) as one 
of the thermodynamic quantities (proportional to the pressure) and thus extended  
the space of macro states. This formalism has attracted considerable interests
because it revealed various critical behaviors and the possible existence of 
different types of phase transitions in black hole physics 
\cite{dolan2010cosmological, dolan2011pressure, dolan2011compressibility, kubizvnak2012p, 
cai2013pv, kubizvnak2017black, xu2014gauss, Xu:2014tja, lemos2018black, 
zhang2015phase,Gunasekaran:2012dq,Poshteh:2013pba,
Altamirano:2013ane,Belhaj:2012bg,Hendi:2012um,Chen:2013ce,
Belhaj:2013cva,Altamirano:2013uqa,Zou:2013owa,Altamirano:2014tva,Zhao:2013oza,
Wei:2014hba,Kubiznak:2014zwa,Zou:2014mha}. 
\blue{There is a plethora of literature pertaining to this topic which seems to be 
impossible to present a full list.  Thus we have only cited the few works that have 
made significant impacts on our line of thinking and researches.}
In particular, the critical phenomena 
of the charged spherically symmetric AdS black holes in conformal gravity, 
whose weird behavior has largely triggered the motivation of the present work, 
is analyzed in \cite{xu2014critical}. 
Further development of the EPS formalism includes the exploration of 
black hole microstructures \cite{wei2015insight,wei2020extended,
dehyadegari2020microstructure}, the inclusion of the central charge and its conjugate
variable in the list of thermodynamic variables \cite{Visser,CKM,Alfaia},
which is inspired by the AdS/CFT correspondence, etc. 

Although the study based on the EPS formalism proves to be very fruitful, it seems that 
several issues are inevitable in this formalism, as first pointed out in 
\cite{gao2021restricted}. To name a few of them, the requirement of a variable 
cosmological constant leads to a theory-changing problem, which we called the 
ensemble of theories problem; the interpretation of the total energy as enthalpy
instead of internal energy seems to be in contradiction with the thermodynamic 
understanding of total energy; some of the thermodynamic behaviors in various 
black hole solutions of different choices of gravity models are so bizarre that 
there is no precedent examples of macroscopic systems with similar behaviors. 
The charged spherically symmetric AdS black hole solution in 
conformal gravity is among the example cases which exhibits very strange behaviors 
in the EPS formalism \cite{xu2014critical}, including the intersecting
isotherms of different temperatures, the discontinuous change of Gibbs free energy in 
thermodynamic processes, the branched or multivalued
thermal equations of states (EOS) and the
appearance of a maximal specific thermodynamic volume (or radius of the event horizon) 
at a fixed temperature, etc. 

Above all, the lack of complete Euler homogeneity
in both the traditional and the EPS formalisms of black hole thermodynamics is
considered to be the most severe problem which constitute a stumbling stone 
in understanding the zeroth law of black hole thermodynamics. 
In order to solve the problem of Euler homogeneity, we proposed a novel 
formalism for black hole thermodynamics called the restricted phase space (RPS)
formalism by fixing the cosmological constant but including the effective 
number of microscopic degrees of freedom and its conjugate chemical potential
in the list of thermodynamic variables. The application of the RPS formalism 
to the cases of RN-AdS \cite{gao2021restricted} and Kerr-AdS \cite{gao2022thermodynamics}
black holes in Einstein gravity indicated that this new formalism is free of all the 
issues mentioned above. Subsequent studies revealed that the RPS formalism works 
for non-AdS black holes as well \cite{wang2021black,zhao2022thermodynamics}, 
and is also applicable to a large class of higher curvature gravity models 
known as black hole scan models \cite{kong2022restricted}. 
According to the behaviors of the charged spherically symmetric AdS black holes in
different black hole scan models, it appears that these models can be subdivided into two 
universality classes, i.e. the Einstein-Hilbert-Born-Infield class and the 
Chern-Simons class. Most recently, we found that the high temperature limit 
of the $(D+2)$-dimension Tangherlini-AdS black holes can be precisely matched to the
low temperature limit of the quantum phonon gases that appear in nonmetallic crystals 
residing in $D$-dimensional flat space \cite{hightemp}, and this 
AdS/phonon gas correspondence seems to be not limited to the cases of Tangherlini-AdS 
black holes, because the high temperature limit of the heat capacities of 
the charged AdS black holes in black hole scan models also  
behave similarly to the low temperature Debye heat capacities of the $D$-dimensional 
quantum phonon gases \cite{kong2022restricted}. 
More applications of the RPS formalism can be found in 
\cite{Sadeghi2022,Bai,Du2022,Wc2022}.

The aim of the present work is two-folded. First we wish to show the applicability 
of the RPS formalism to the case of four dimensional conformal gravity, and resolve 
the bizarreness in the thermodynamic behaviors of the charged AdS black holes in this 
model that appeared in the EPS formalism. Second, we wish to take conformal gravity 
as yet another example case for the AdS/phonon gas correspondence. As will be clear 
in the main text, these two-folded aims are perfectly accomplished.

\section{The model and the solution}

The model which we consider in this work is best described by its classical action
\begin{align}
S=\alpha\int \rd^4x \sqrt{-g}\left(\frac{1}{2}C^{\mu\nu\rho\sigma}
C_{\mu\nu\rho\sigma} +\frac{1}{3} F^{\mu\nu}F_{\mu\nu}\right),
\label{action}
\end{align}
where the unusual sign in front of the Maxwell term is inspired by critical
gravity \cite{Lu:2011zk} and is necessary for the Einstein
gravity to emerge in the infrared limit
\cite{Maldacena:2011mk}. 

The static charged AdS black hole solution for this model 
is found in \cite{Li:2012gh}, with the metric
\begin{align}
&\mathrm{d} s^2=-f(r)\mathrm{d}t^2+\frac{\mathrm{d}r^2}{f(r)}+r^2\mathrm{d}
\Omega_{2,\epsilon}^2, \label{metric}\\
&f(r)=-\frac{1}{3}\Lambda r^2+c_1 r+c_0+\frac{d}{r},\label{fr}
\end{align}
and the Maxwell field
\begin{align}
A=-\frac{Q}{r}\mathrm{d}t.
\end{align}
The parameter $\epsilon$ can take three discrete values $-1, 0, 1$ which correspond,
respectively, to the hyperbolic, planar and spherical geometry of the 
2-dimensional ``internal space'' characterized by the line element 
$\mathrm{d}\Omega_{2,\epsilon}^2$. The solutions with $\epsilon=0,-1$ 
are known as topological black holes, which exist only in AdS backgrounds. 
The other parameters $Q, c_0, c_1, d, \Lambda$ 
are all integration constant which need to obey an additional constraint
\begin{align}
  3c_1d+\epsilon^2+Q^2=c_0^2.
\label{relation}
\end{align}

The above solution describes a charged AdS black hole provided $\Lambda<0$ and 
the equation
\begin{align}
f(r_0)=0
\label{evh}
\end{align}
has a nonvanishing real positive root $r_0$ which corresponds to the radius of the event
horizon. Notice that the function
$f(r)$ does not contain a term $Q^2/r^2$ as in the usual RN-AdS black hole solution. 
The parameter $Q$ (related to the electric charge) affects the geometry of the spacetime 
only implicitly through the constraint condition \eqref{relation}. Since $Q$ 
appears only in squared form in eq.\eqref{relation}, the spacetime geometry 
does not discriminate positive and negative values of $Q$. Therefore, in this paper, we will
consider exclusively the choice $Q\geq0$. The opposite choice is permitted but 
makes no difference regarding the geometry and thermodynamic behaviors.

In the absence of the parameter $c_1$ (which represents a massive spin-2 hair
\cite{Li:2012gh,Lu:2012xu}), 
the metric \eqref{metric} looks very similar to that of the standard Schwarzschild-(A)dS 
black hole, provided $d$ takes a negative value. Let us stress that, 
unlike in most of the other gravity models, the cosmological constant $\Lambda$ 
arises purely as an integration constant, therefore, this model was once considered to be 
very appropriate for pursuing thermodynamic analysis following the extended phase space 
approach, because a variable cosmological constant in this 
model does not cause the ensemble of theories problem which appears in other 
theories of gravity. 

\section{EPS thermodynamics revisited}

Before rushing into the RPS formalism for the thermodynamics
of the above black hole solution, let us first make a brief review on the 
EPS description and point out some of its pathologies.

To begin with, let us present the relevant thermodynamic quantities in the EPS formalism. 
First, the total energy $E$ of the black hole spacetime, as calculated using 
the Noether charge associated with the
timelike Killing vector \cite{Lu:2012xu}, reads
\begin{align}
E&=\frac{\alpha\omega_2}{24\pi}\left[
\frac{(c_0-\epsilon)(\Lambda r_0^2-3c_0)}{3 r_0}
  +\frac{(2\Lambda r_0^2-c_0+\epsilon)d}{r_0^2}\right],
\label{Evalue}
\end{align}
which is regarded to be the enthalpy, where $r_0 > 0$ represents the radius 
of the event horizon of the black hole and is a real root of $f(r)$, and
$\omega_2$ is the volume of the internal $2d$ space designated by the line element 
$\mathrm{d}\Omega_{2,\epsilon}^2$. The pressure $P$ and its conjugate, 
the thermodynamic volume $V$, are given respectively as
\begin{align*}
  P=-\frac{\Lambda}{8\pi},\qquad
  V=-\,{\frac {\alpha\omega_2d}{3 }}.
\end{align*}
Next comes the black hole temperature and entropy, which are given by \cite{Li:2012gh}
\begin{align}
T&=-\,{\frac {\Lambda\,r_0^{3}+3c_{0}r_{0}+6d}{12\pi{r_{0}}^{2}}},
\quad
S=\,{\frac {\alpha\omega_2\left(\epsilon r_0-c_{0}r_{0}-3d \right) }{6r_{0}}}.
\label{temperature}
\end{align}
The electric charge (defined as the conserved charge associated with the $U(1)$ 
gauge symmetry of the electromagnetic field) and the conjugate potential are 
given respectively
\begin{align}
&Q_{{e}}=\,{\frac {\alpha\omega_2Q}{12\pi }},\qquad \Phi=-{\frac {Q}{r_{0}}}.
\label{charge}
\end{align}
The parameter $c_1$ is a massive spin-2 hair which is also taken as one of the 
thermodynamic variables. This variable is denoted as $\Xi=c_1$, and its conjugate 
$\Psi$ is given by
\begin{align*}
\Psi=\,{\frac {\alpha\omega_2\left( c_{0}-\epsilon \right) }{24\pi }}.
\end{align*}
In the above expressions for thermodynamic quantities, the parameter $c_0$ is 
considered to be implicitly determined via the relation \eqref{relation}, 
and thus it cannot be taken as a simple constant while considering 
thermodynamic behaviors. On the contrary, the coupling constant $\alpha$ is 
always kept as a real constant in the EPS formalism.

It can be checked that the energy $E$ obeys the following relations,
\begin{align*}
{\rd E}&=T{\rd S}+\Phi\,{\rd Q}_{{e}}+ \Psi\,\rd \Xi + V\,\rd P,\\
E&=2PV+\Psi\,\Xi. 
\end{align*}
These relations are interpreted as the first law and the Smarr relation in the 
EPS formalism. 

As mentioned in the introduction, the behavior of the 
EPS thermodynamics as outlined above appears to be very strange, and 
the problems may be attributed either to the gravity model itself
or to the EPS formalism. We will see that, with new 
insights from the RPS formalism, the problems can be 
perfectly avoided, therefore, it is clear that the problems arise from the EPS 
formalism.

\section{The RPS formalism}

The whole logic of the RPS formalism stands as follows. First of all, we need to introduce 
a new pair of thermodynamic variables, i.e. the effective number $N$ 
of microscopic degrees of freedom (\blue{or dubbed {\em black hole molecules}}) 
of the black hole and its conjugate, the chemical 
potential $\mu$. These two objects are universally defined as
\begin{align}
N=\frac{L^D}{G},\qquad  \mu=\frac{GT I_E}{L^D},
\label{Nmu}
\end{align}
where $L$ is an arbitrarily chosen constant length scale, $G$ is the 
Newton constant, and $I_E$ is the on-shell Euclidean action which corresponds to 
the black hole solution. \blue{The arbitrariness of $L$ may be attributed to the fact that 
we do not actually know what a black hole molecule is, but this does not prevent us from 
describing the macroscopic properties of the black hole, just like in the studies of  
thermodynamics of ordinary matter systems in which the precise nature of
individual molecules does not matter, and the total number of molecules can be 
taken as an arbitrary number as long as the whole system remains macroscopic, 
which means that $L$ should be sufficiently large in our present case.}

In the present case, one has $D=2$ and
\cite{Li:2012gh},
\begin{align}
I_E=\frac{\alpha\omega_2\left[2(c_0-\epsilon)\epsilon r_0+(3\epsilon-\Lambda r_0^2)d\right]}
{24\pi r_0^2 T}.
\label{IE}
\end{align}
Since the Newton constant does not explicitly appear in the action \eqref{action}, 
we need to relate the coupling constant $\alpha$ to $G$ in some way. Due to the fact that
$G$ has dimension $[L]^2$, while $\alpha$ has dimension $[L]^0$, This relationship 
could not be the naive choice $\alpha=1/16\pi G$ but rather needs to be modified 
by a factor of dimension $[L]^2$. Therefore, we assume that
\begin{align}
\alpha=\frac{L^2}{16\pi G}, \label{aG}
\end{align}
thanks to the arbitrariness of $L$. \blue{This assumption is not an absolutely necessary 
step. What actually matters is that the number of black hole molecules 
$N$ should be proportional to the overall factor in the action, in order to ensure that $N$ 
is the thermodynamic conjugate of the chemical potential $\mu$ (which in turn is defined 
in terms of the Euclidean action $I_E$). The assumption \eqref{aG} is introduced 
simply for illustrating that the overall factor $1/16\pi G$ in 
Einstein gravity and the factor $\alpha$ in conformal gravity play similar roles 
in the RPS formalism.}

Before checking the thermodynamic relations in the RPS formalism, there is something 
more to be fixed in the solution \eqref{metric}, \eqref{fr}. In order to guarantee 
the existence of a reasonable weak field limit with attractive Newtonian potential, 
we need to require that
\begin{align}
1/|c_1|\gg r_0, \quad 1/\sqrt{-\Lambda}\gg r_0, \quad d<0.
\end{align}
Moreover, in the extremal case with $c_1=d=0$, the solution must fall back to that of
the vacuum AdS background. Therefore, for each choice of $\epsilon=-1,0,1$, 
$c_0$ must always be equal to $\epsilon$. 

Let us remark that, in previous studies, $c_0$ was considered to be an
implicit function in $c_1, d$ and $Q$. The present choice $c_0=\epsilon$ 
is more physically motivated, which makes a big difference. In fact, if 
$c_0$ and $\epsilon$ were kept independent besides the constraint \eqref{relation}, 
the first law in the RPS formalism
to be introduced below would not hold.

In the following, we shall be working exclusively with the choice $c_0=\epsilon=1$, 
which corresponds to the spherically symmetric case with $\omega_2=4\pi$. Inserting 
$c_0=\epsilon=1$, $\omega_2=4\pi$ and eq.\eqref{aG} into eqs.\eqref{Evalue}, 
\eqref{temperature}, \eqref{charge} and eqs.\eqref{Nmu}, \eqref{IE}, we get
\begin{align}
E&=\frac{\Lambda L^2 d }{48\pi G}, \label{E1}\\
S&=-{\frac {L^2 d }{8 G r_{0}}}, \qquad \quad\,\,\,
T=-\frac {r_0(\Lambda\,r_0^{2}-3)+6(d+r_{0})}{12\pi{r_{0}}^{2}}, \label{TS1}
\\
Q_{{e}}&=\,{\frac {L^2 Q}{48\pi G }},\qquad \qquad \Phi=-{\frac {Q}{r_{0}}},
\label{QPhi1}\\
N&=\frac{L^2}{G}=\blue{16\pi\alpha},\qquad
\mu= -\frac{(\Lambda r_0^2-3)d}{96\pi r_0^2}.
\label{Nmu1}
\end{align}
It is now more transparent that, in order to ensure non-negativity of 
the entropy and the total energy, $d$ and $\Lambda$ must be both negative. 
That is why the solution is considered to be an AdS black hole solution 
from the very beginning. 

In the above equations, $d, Q, G$ and $r_0$ are considered to be implicit functions in the 
thermodynamic variables. However, these objects are not all independent due 
to the constraint condition \eqref{relation} and the equation for the event horizon 
\eqref{evh}. The joint system of equations \eqref{relation} and \eqref{evh} has two 
different sets of solutions for $d$ and $c_1$, among which only one ensures negativity 
of $d$, 
\begin{align}
d&= \frac{r_0}{6}\left[\Lambda r_0^2-3 - \sqrt{(\Lambda r_0^2-3)^2+12Q^2}\right],\\
c_1&=\frac{1}{6r_0} \left[\Lambda r_0^2-3 + \sqrt{(\Lambda r_0^2-3)^2+12Q^2}\right].
\end{align}
\blue{The other solution with reversed signs in front of the square roots corresponds to
strictly non-negative values of $d$, which results in a non-positive entropy, and 
therefore will be dropped henceforth.}

\blue{Notice that, in the RPS formalism, we do not introduce the $(\Xi,\Psi)$ variables. 
The reason behind this choice is the symmetry principle. Let us quote 
C.N. Yang's celebrated dictum: ``{\em Symmetry dictates dynamics}.''  
Here we would like to extend this statement a little step further: 
{\em Symmetry dictates thermodynamics}, which means that different thermodynamic 
systems with the same underlying symmetries should be described by the same set 
of thermodynamic variables, although their detailed thermodynamic behaviors could differ 
from each other. The black hole solution under study bears the same spherical symmetry 
and the same $U(1)$ gauge symmetry as the well-known RN black hole solution in 
Einstein gravity, thus the space of macro states for these two black hole systems need to be  
spanned by the same set of macroscopic variables. This may help for understanding why 
we exclude the variables $(\Xi, \Psi)$ from the list of allowed thermodynamic quantities
in the RPS formalism. Let us stress that, {\em it is the underlying symmetries, 
rather than the 
number of integration constants, that determine the dimension of the space of macro states.}
One may wonder why the conformal symmetry is not taken into account
in our consideration. The reason is quite clear: any concrete choice of metric in 
conformal gravity automatically breaks the conformal symmetry. Therefore, no charges 
associated with the conformal symmetry could enter into the thermodynamic description 
for black holes in conformal gravity.
}

Using the results presented in eqs.\eqref{E1}-\eqref{Nmu1}, 
one can check by straightforward calculations that the first law
\begin{align}
\rd E= T\rd S+\Phi\rd Q_e +\mu\rd N
\end{align}
and the Euler relation
\begin{align}
E=TS+\Phi Q_e+\mu N \label{Eu}
\end{align}
hold simultaneously, which ensures that $E$ is a first order homogeneous function in 
$(S, Q_e, N)$ and that $T, \Phi, \mu$ are zeroth order homogeneous functions in 
$(S, Q_e, N)$. The last two equations
constitute the fundamental relations for the RPS formalism
of black hole thermodynamics. As a direct consequence, the Gibbs-Duhem relation
\begin{align}
\rd \mu =-s\rd T-q \rd\Phi 
\end{align}
also holds, where 
\[
s=\frac{S}{N}=-\frac{d}{8r_0}, \quad
q=\frac{Q_e}{N}=\frac{Q}{48\pi}, 
\]
both are zeroth order homogeneous functions in $(S, Q_e, N)$.
The Gibbs-Duhem relation indicates that 
the intensive variables $\mu, T, \Phi$ are not independent of each other, whereas each 
of them is independent on the size of the black hole.

\section{Thermodynamic processes in the RPS formalism}

The explicit values of the various thermodynamic quantities collected in 
the last section allow for a detailed analysis on the thermodynamic behavior of the 
black hole solution under consideration. To proceed, we first need to re-express
the parameters $r_0, G, Q$ in terms of the extensive variables $N, S, Q_e$
or better in terms of $N$ and $s,q$. The latter set of variables has the advantage that 
the EOS re-expressed in these variables are independent of $N$, 
which is a characteristic property of standard extensive thermodynamic systems
known as the law of corresponding states.

In order to rewrite $r_0, G, Q$ as functions in $(N,s,q)$, we need to solve 
the first equations in eqs.\eqref{TS1}-\eqref{Nmu1} as a system of algebraic 
equations for $r_0, G, Q$, which yields
\begin{align}
Q=48\pi q,\quad G=\frac{L^2}{N},\quad
r_0=\ell s^{-1/2}(8s^2-s-96\pi^2q^2)^{1/2},
\label{para}
\end{align}
where $\ell\equiv\nbfrac{3}{\Lambda}^{1/2}>0$. The condition for $r_0$ to be
real and positive reads
\begin{align}
8s^2-s-96\pi^2q^2>0.    \label{Scond}
\end{align}
Since $q^2\geq0$ and $s>0$ (a macro state of zero entropy could not be understood 
as a black hole), we can deduce from the above inequality that $s>1/8$. In other words,
there is no black hole states with $s\leq 1/8$.

There is another, negative-valued, unphysical, solution for $r_0$ which is omitted.

Inserting eq.\eqref{para} into eq.\eqref{E1} and the rest equations in 
eqs.\eqref{TS1}-\eqref{Nmu1}, we have
\begin{align}
\mathcal{E} &= s^{1/2} (8s^2-s-96\pi^2q^2)^{1/2}, \label{Ee}\\
\tau &= \frac{12s^2-s-48\pi^2q^2}{s^{1/2}(8s^2-s-96\pi^2q^2)^{1/2}},
\label{eostau} \\
\phi &= -\frac{96\pi^2 q s^{1/2}}{(8s^2-s-96\pi^2q^2)^{1/2}},
\label{eosphi} \\
m &= -\frac{4s^{1/2}(s^2-12\pi^2q^2)}{(8s^2-s-96\pi^2q^2)^{1/2}}, \label{m}
\end{align}
where, for convenience, we introduced the new variables
\begin{align}
\mathcal{E}=\frac{2\pi\ell}{N} E,\quad 
\tau= 2\pi\ell T,\quad 
\phi= 2\pi\ell \Phi,\quad
m= 2\pi\ell \mu,
\end{align}
each has dimension $[L]^0$. The first order homogeneity of $E$ and zeroth order 
homogeneity of $T,\Phi, \mu$ are transparent in the above expressions.
Notice that the condition \eqref{Scond} automatically ensures 
the non-negativity of $T$, therefore, there is no further constraints over the 
parameters $s$ and $q$. 

Besides the above thermodynamic quantities, we also need the explicit expression 
for the Helmholtz free energy $F=E-TS$, which, in terms of the rescaled variable 
$f\equiv 2\pi\ell F/N$, is given as follows,
\begin{align}
f=\mathcal{E}-\tau s
=- \frac{4s^{1/2}(s^2+12\pi^2q^2)}{(8s^2-s-96\pi^2q^2)^{1/2}}.
\label{Ff}
\end{align}

The first thing to be noticed from the above results is the absence of Hawking-Page 
(HP) transition in the present case. The HP
transition is a phase transition from an AdS black hole to a pure thermal gas 
which occurs at the zero of the Gibbs free energy (or equivalently of the chemical 
potential) for neutral AdS black holes \cite{Hawking2}. 
Such transition is known to exist in most AdS black hole solutions in 4 and higher 
spacetime dimensions. However, in the present case, the chemical potential is strictly 
negative in the neutral limit, as can be inferred from by eq.\eqref{m}. 
This seems to indicate that the AdS black hole solution in conformal gravity belongs 
to a novel universality class which is different from the classes of 
AdS black holes either in the Einstein-Hilbert and Born-Infield like theories or in 
the Chern-Simons like theories. 

The concrete thermodynamic behaviors of the black hole can be graphically illustrated by 
plotting the EOS \eqref{eostau}-\eqref{m}. 

\begin{figure}[ht]
\begin{center}
\includegraphics[width=.42\textwidth,height=.22\textheight]{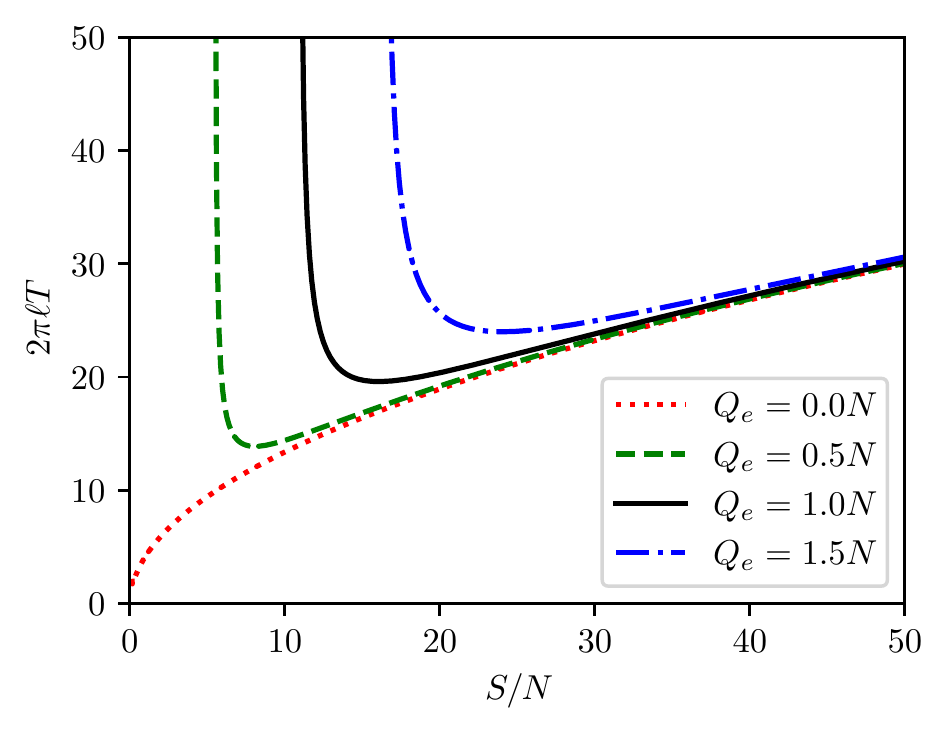} \includegraphics[width=.44\textwidth,height=.22\textheight]{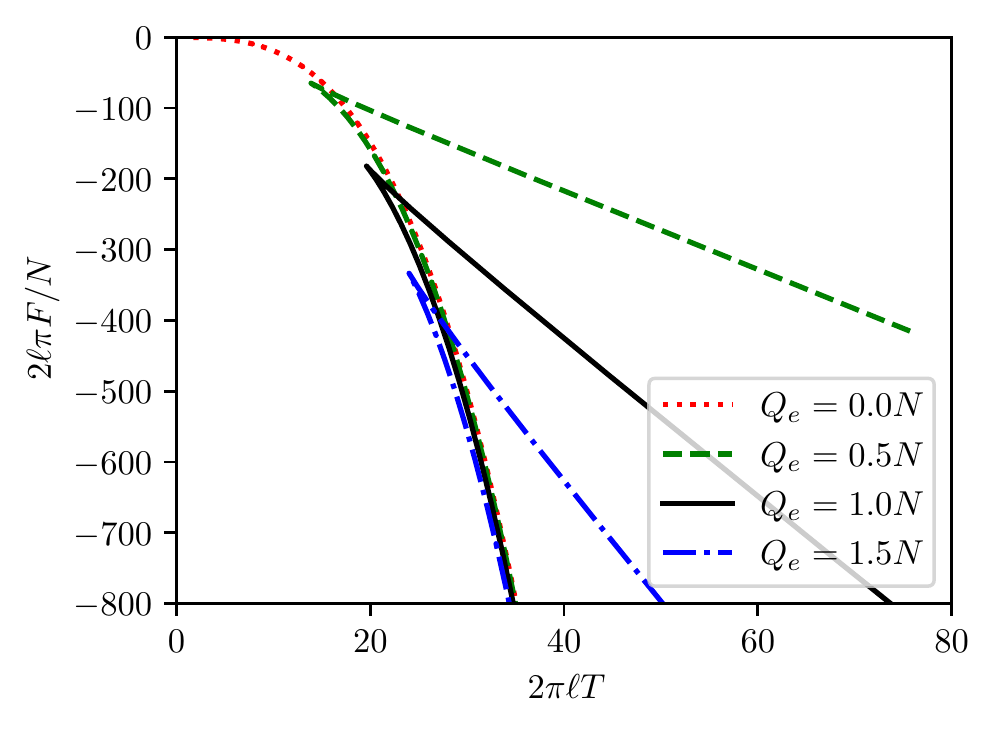}

\caption{The isocharge $T-S$ and $F-T$ curves}\label{fig1}
\end{center}
\end{figure}

First let us look at the isocharge $T-S$ and $F-T$ curves presented in Fig.\ref{fig1}.
Each isocharge $T-S$ curve contains a single minimum which divides 
the black hole states of the same temperature and the same charge into two branches,
i.e. unstable small black hole and stable large black hole. 
Correspondingly, the $F-T$ curves are also branched, with the lower branch 
corresponding to the stable large black hole states. 
Above the minimal temperature, the transition from the unstable small black hole state
to the large stable black hole state should take place under small perturbations. 
There is no equilibrium condition for such transitions. 

\begin{figure}[ht]
\begin{center}
\includegraphics[width=.42\textwidth,height=.22\textheight]{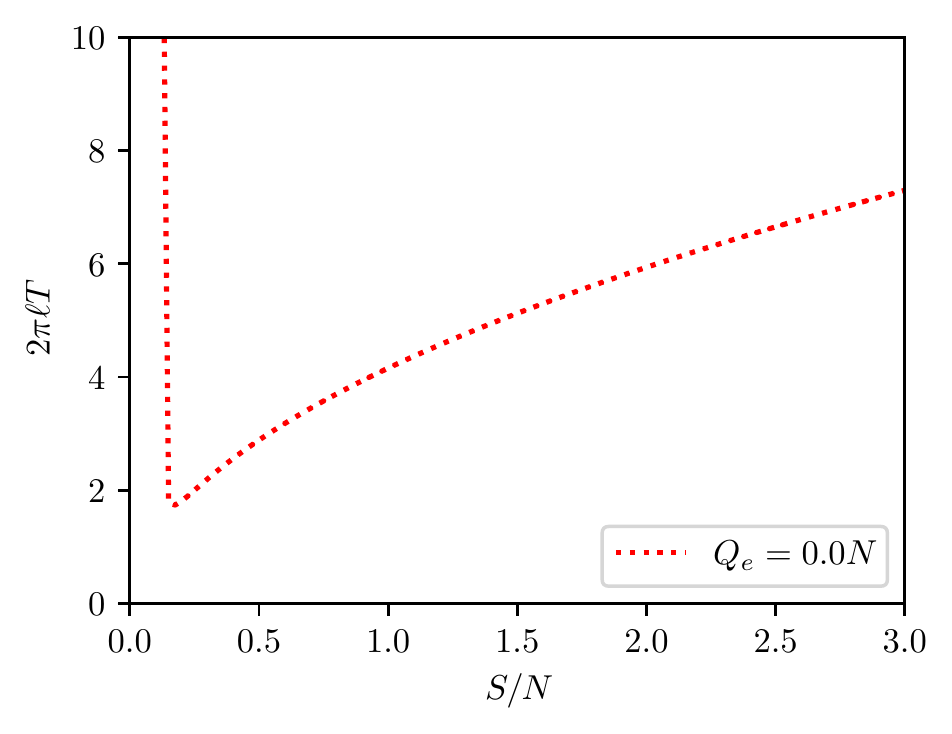} \includegraphics[width=.44\textwidth,height=.22\textheight]{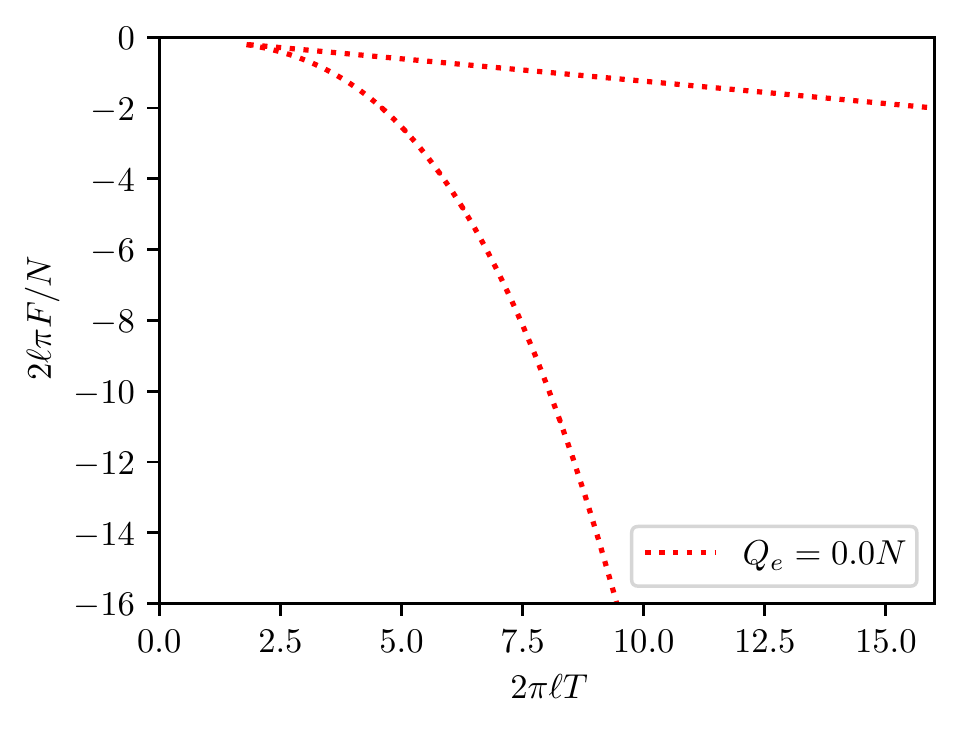}

\caption{Zoomed-in plots of the isocharge $T-S$ and $F-T$ curves}\label{TS0}
\end{center}
\end{figure}

From the curves depicted in Fig.\ref{fig1} it appears that the description 
on the behavior of the isocharge $T-S$ and $F-T$ curves presented in the 
last paragraph does not work in the case with $Q=0$. 
This is not true. Fig.\ref{TS0} presents the zoomed-in plots for 
the isocharge $T-S$ and $F-T$ curves near the origin. 
It can be seen that the above branched behavior persists at $Q=0$.

The above isocharge $T-S$ behavior
is in sharp contrast to the case of charged AdS black holes in Einstein-Hilbert, 
Born-Infield or Chern-Simons like theories of gravity. In the case of  
Einstein-Hilbert and Born-Infield like theories, the isocharge $T-S$ processes 
always contain an equilibrium phase transition which is of the first order
above the critical temperature and becomes second order at the critical point, 
while in the case of Chern-Simons like theories, the isocharge $T-S$ curves 
are monotonic and there is only one stable black hole state at each fixed 
temperature and electric charge. It looks that the present model gives a third 
universality class which interpolate the 
Einstein-Hilbert-Born-Infield class and the Chern-Simons class. 

\begin{figure}[ht]
\begin{center}
\includegraphics[width=.42\textwidth,height=.23\textheight]{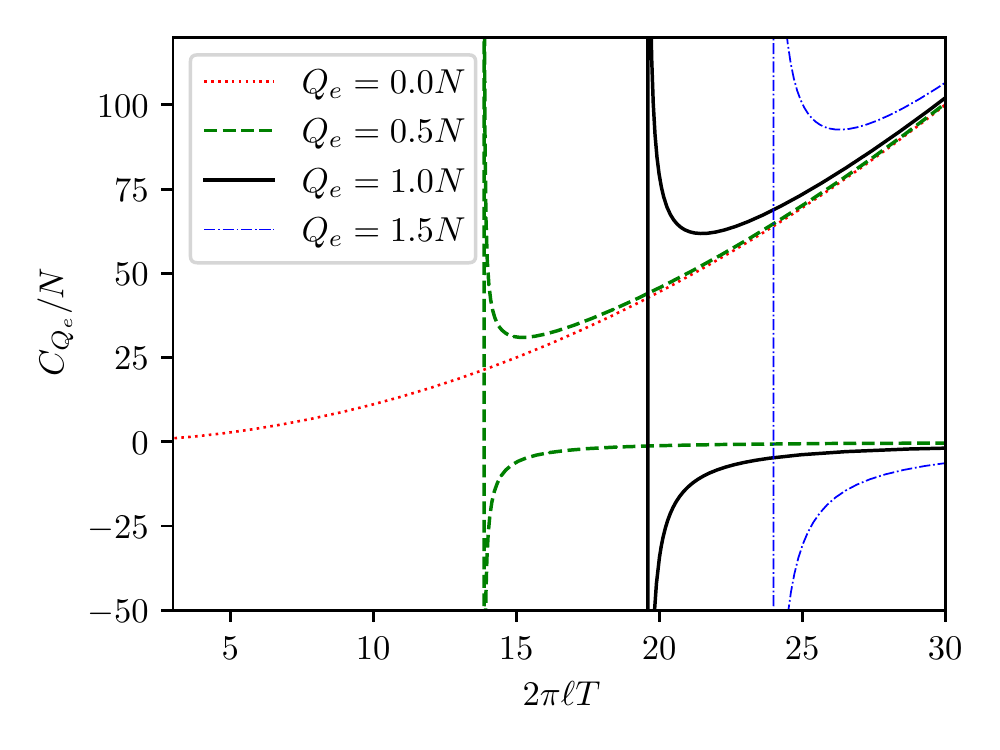} 
\includegraphics[width=.45\textwidth,height=.23\textheight]{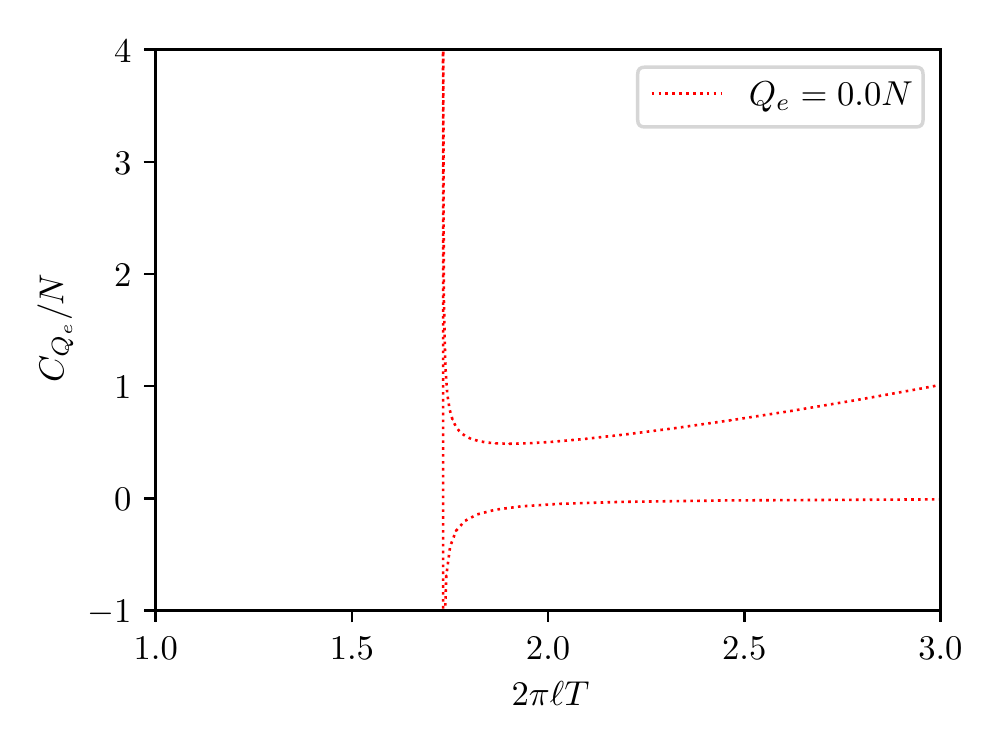}

\caption{The isocharge heat capacity versus temperature}\label{fig2}
\end{center}
\end{figure}

The branched behavior of the isocharge processes can also be revealed from the 
behavior of the isocharge heat capacity $C_{Q_e}$. In the present case, 
the isocharge specific heat capacity $c_q\equiv C_{Q_e}/N$ 
can be calculated explicitly using the EOS 
\eqref{eostau},
\begin{align}
c_q &= T \pfrac{s}{T}_q =\frac{\tau}{\pfrac{\tau}{s}_q}\nonumber\\
&=-\frac{s \left[s (8 s-1)-96 \pi ^2 q^2\right] \left[s (12 s-1)-48 \pi ^2
   q^2\right]}{8 \left(288 \pi ^4 q^4+144 \pi ^2 q^2 s^2-6 s^4+s^3\right)}.
\end{align}
Based on this result, the isocharge heat capacity versus temperature curves
are plotted in Fig.\ref{fig2}, wherein the right figure is the zoomed-in 
plot of the curve with $Q_e=0$. The branched behavior is transparent, and 
only the large black hole branch has positive heat capacity 
which indicate its stability.

\begin{figure}[ht]
\begin{center}
\includegraphics[width=.42\textwidth,height=.23\textheight]{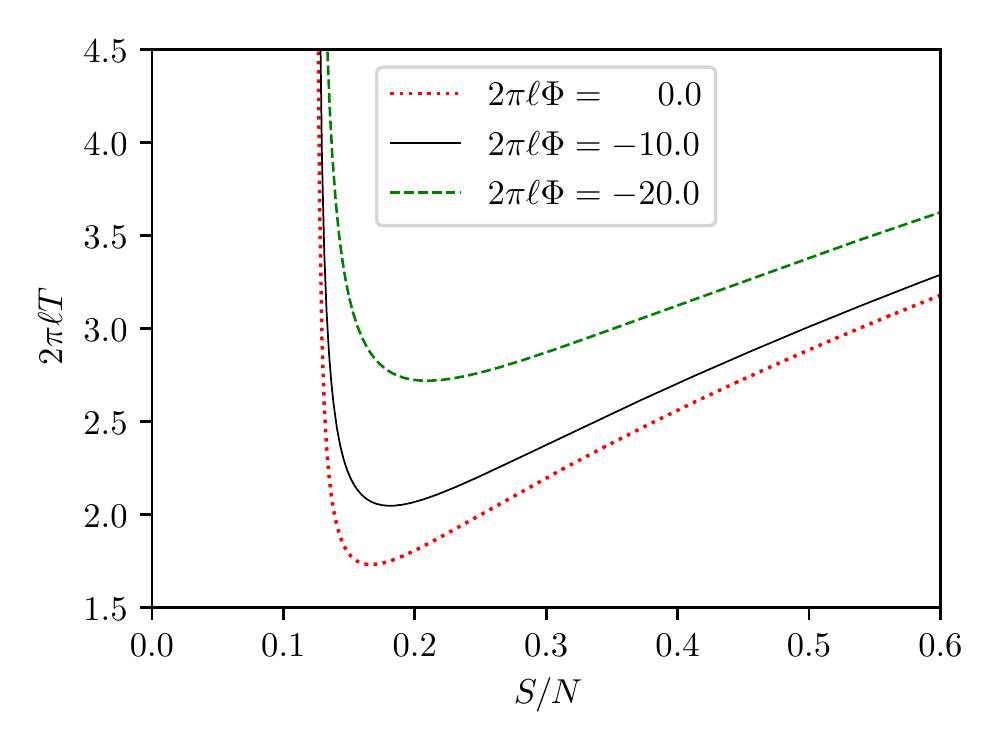}
\includegraphics[width=.42\textwidth,height=.23\textheight]{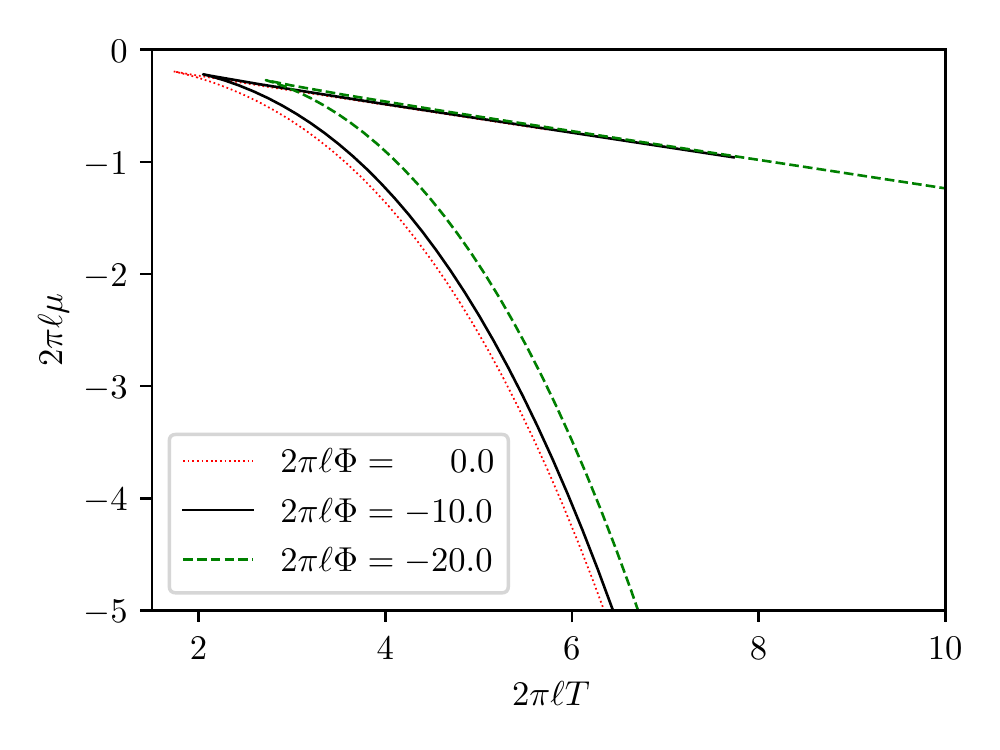}

\caption{The isovoltage $T-S$ and $\mu-T$ curves}\label{fig3}
\end{center}
\end{figure}

The branched behavior also appears in the isovoltage $T-S$ processes, as 
depicted in Fig.\ref{fig3} together with the isovoltage $\mu-T$ curves. 
The isovoltage $T-S$ curves are qualitatively similar to
that of the charged AdS black holes in Einstein-Hilbert-Born-Infield class of theories. 
However, in the isovoltage processes, $\mu$ is strictly negative for any choice of $T$, 
which is consistent with the earlier statement on the absence of HP transition. 

\begin{figure}[ht]
\begin{center}
\includegraphics[width=.42\textwidth,height=.23\textheight]{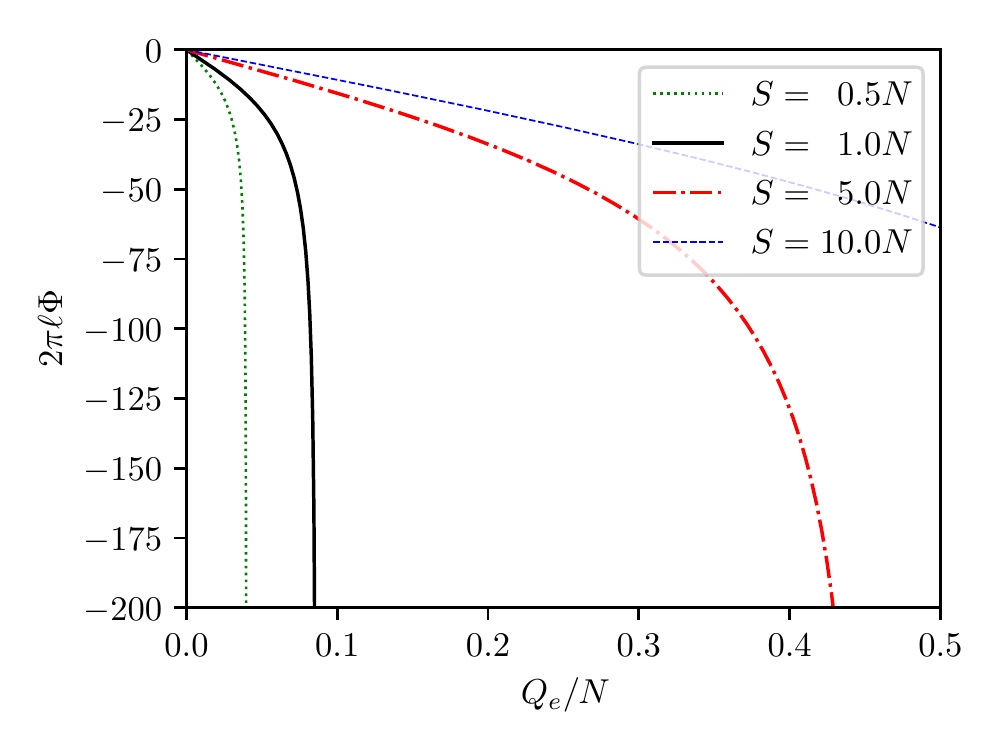}
\includegraphics[width=.42\textwidth,height=.23\textheight]{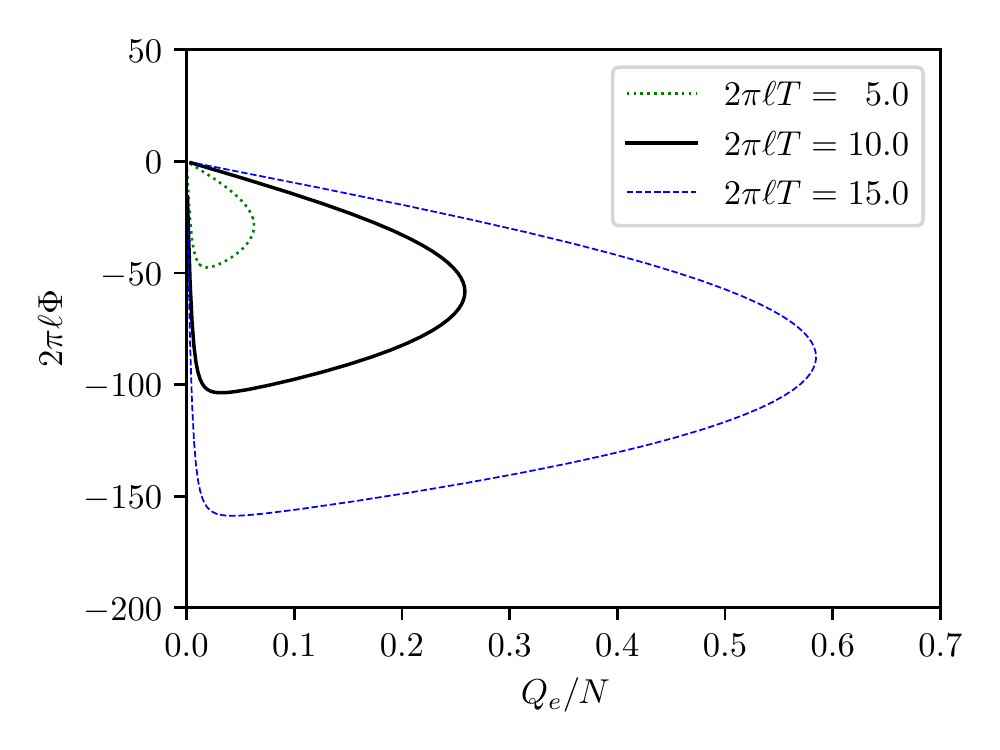}

\caption{The adiabatic and isothermal $\Phi-Q_e$ curves}\label{PhiQ}
\end{center}
\end{figure}

Besides the $T-S$ processes, one may also be interested in the $\Phi-Q_e$ processes. 
There are two possible types of $\Phi-Q_e$ processes, i.e. adiabatic and isothermal. 
The corresponding  $\Phi-Q_e$ curves are depicted in Fig.\ref{PhiQ}. In the 
adiabatic processes, the electric potential decreases monotonically as the charge 
increases, and there is an upper bound for $Q_e$ at each fixed $S$ as can be seen in
eq.\eqref{Scond}. The isothermal $\Phi-Q_e$ processes appear to be more involved. 
Besides the existence of an upper bound for $Q_e$ at each fixed $T$, it seems that, 
at each fixed $T$, the black hole could experience a cyclic charging-discharging process, 
which makes the black hole as a potential battery. This kind of charge-potential 
process has not been found previously within the RPS formalism for the thermodynamics
of charged AdS black holes.

\begin{figure}[ht]
\begin{center}
\includegraphics[width=.42\textwidth,height=.23\textheight]{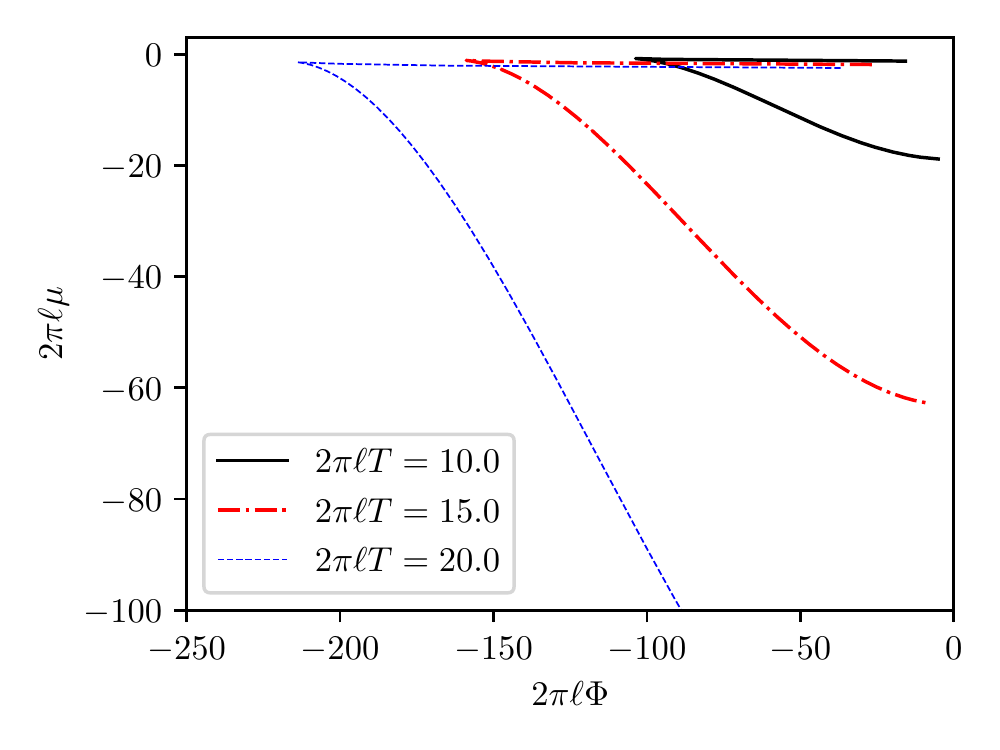}

\caption{The isothermal $\mu-T$ curves}\label{muphi}
\end{center}
\end{figure}

One may also look at the isothermal $\mu-\Phi$ curves
presented in Fig.\ref{muphi}. As expected, the isothermal $\mu-\Phi$ curves
also possess a branched behavior at each fixed $T$, with the lower branch 
being the stable large black hole branch.

\blue{As a final remark, let us consider the special case $c_{1}=Q=0$. 
The metric function $f(r)$ now becomes
\[
f(r)=1+\dfrac{d}{r}-\dfrac{1}{3}\Lambda r^{2}.
\]
With negative $d$, the corresponding metric takes the same form as that of the 4d
Schwarzschild-AdS black hole solution in Einstein gravity. However, this similarity 
does not imply that the thermodynamic behaviors should also degenerate to that of the 
4d Schwarzschild-AdS black hole solution in Einstein gravity, because only part of the 
thermodynamic quantities (e.g. the temperature $T$ and the electric potential $\Phi$) 
of the black hole is determined solely by the solution, whilst the other part of the 
thermodynamic quantities , including 
the energy $E$, the entropy $S$, the charge $Q_e$ and the chemical potential $\mu$, 
is actually determined by the action of the underlying gravity model. Thus the same 
metric as solution of different gravity models does not necessarily have the 
same thermodynamic behavior. This final statement is justified by the absence of HP 
transition in the case of conformal gravity as illustrated in Fig.\ref{fig3} and 
the presence of HP transition in the case of neutral limit of RN-AdS black holes in 
Einstein gravity \cite{gao2021restricted}.
}

\section{High temperature limit in the stable branch}

Our recent study \cite{hightemp} on the case of Tangherlini-AdS black hole solution 
in Einstein gravity revealed a remarkable 
connection between the high temperature limit of AdS black holes and the low 
temperature limit of phonon gases in nonmetallic crystals. It is natural to test whether
this AdS/phonon gas correspondence still holds in the case of charged spherically 
symmetric AdS black holes in conformal gravity. 

Please be reminded that the AdS/phonon gas correspondence 
revealed in \cite{hightemp} holds only in the stable large black hole branch. 
Therefore, we also consider the high temperature limit of the stable large black hole 
branch. Using the results presented in eqs.\eqref{Ee}-\eqref{Ff}, 
we can easily get
\begin{align}
&\lim_{s\to\infty} \frac{\mathcal E}{\tau^3} = \frac{1}{27},\quad
\lim_{s\to\infty} \frac{f}{\tau^3} = -\frac{1}{54},\\
&\lim_{s\to\infty} \frac{s}{\tau^2} = \frac{1}{18},\quad
\lim_{s\to\infty} \frac{c_q}{\tau^2} = \frac{1}{9}.
\end{align}
In the stable large black hole branch, $s\to\infty$ implies $T\to\infty$. Therefore, the 
above limits can be easily translated into the following high temperature limits for the
thermodynamic quantities,
\begin{align}
\lim_{T\to\infty} E &= \frac{1}{27}(2\pi\ell)^2 N T^3,\quad
\lim_{T\to\infty} F = -\frac{1}{54}(2\pi\ell)^2 N T^3,\\
\lim_{T\to\infty} S &= \frac{1}{18} (2\pi\ell)^2 N T^2,\quad
\lim_{T\to\infty} C_{Q_e} = \frac{1}{9}(2\pi\ell)^2 N T^2.
\end{align}
The high temperature limit means that the physical temperature is high above 
some constant characteristic temperature. In the present case, the characteristic 
temperature can be chosen as 
\begin{align}
T_{\rm bh} = \frac{3\sqrt{2}}{2\pi\ell}.
\end{align}
With this choice, the above high temperature behaviors can be rewritten as 
\begin{align}
E&\approx \frac{2}{3}NT\bfrac{T}{T_{\rm bh}}^2,\quad\,\,\,
F\approx -\frac{1}{3}NT\bfrac{T}{T_{\rm bh}}^2,\\
S&\approx N\bfrac{T}{T_{\rm bh}}^2,\quad\quad
C_{Q_e}\approx 2NT\bfrac{T}{T_{\rm bh}}^2
\end{align}
under the condition $T\gg T_{\rm bh}$. These behaviors coincide precisely with what 
we obtained for Tangherlini-AdS black holes in generic spacetime dimensions 
$D+2\geq 4$, even the numerical coefficients are the same if we set $D=2$. 
As we have already pointed out in \cite{hightemp}, such high temperature 
asymptotic behaviors also 
coincide with the low temperature behaviors of the quantum phonon gases in 
nonmetallic crystals. By the way, we also checked the 
cases of four dimensional Kerr-AdS and Kerr-Newman-AdS black holes and found that
the high temperature asymptotic behaviors of the above thermodynamic quantities are 
exactly the same. For different choices of AdS black holes, the high temperature asymptotic
behaviors differ from each other at most by the choice of different characteristic 
temperatures.

\section{Concluding remarks}

The thermodynamics of charged, spherically symmetric, AdS black holes in 
four dimensional conformal gravity theory is reconsidered using the RPS formalism. 
The strange thermodynamic behaviors found previously 
within the EPS formalism completely disappear, including the strange 
multivalued and intersecting isotherms, and the zeroth order phase transitions 
with discontinuities in the Gibbs free energies. 
Instead, the complete Euler homogeneity, which is known to be absent in the EPS formalism,  
is restored in the RPS formalism. Therefore, the results of the RPS formalism looks 
simpler and is physically more reasonable.

Detailed study on the thermodynamic processes seems to indicate that the RPS thermodynamics 
of charged spherically symmetric AdS black holes in conformal gravity theory may 
belong to a brand new universality class as opposed to the classes of charged 
spherically symmetric AdS black holes in 
Einstein-Hilbert/Born-Infield like theories and in Chern-Simons like theories of gravity.

Let us recall that the major difference between the latter two universality classes 
lies in that, in the case of Einstein-Hilbert/Born-Infield like theories, the 
isocharge $T-S$ processes contain a first order supercritical phase transition 
which becomes second order at the critical point, while in the case of 
Chern-Simons like theories, the isocharge $T-S$ processes contain no phase 
transitions at all. The common property of these two classes of theories lies in that, 
in both cases, each of the isovoltage $T-S$ curves contains a single minimum, 
indicating the existence of non-equilibrium and noncritical phase transition from the small
unstable black hole branch to the large stable black hole branch, and that, in the high
temperature limit, the thermodynamic behaviors of the black holes can be precisely 
matched to that of the low temperature limit of the quantum phonon gases residing in 
$D$ dimensional flat space, with $D$ being equal to the dimension of the 
bifurcation horizon of the black holes. Another common feature of the above 
two universality classes is the existence of HP transition in the neutral limit.

The results presented in the present work indicate that, the thermodynamic behavior
of the present model is quite different from the above two universality classes. 
Here we can list three distinguished features of the present case. 
Firstly, each of isocharge $T-S$ curves contains 
a single minimum just like the isovoltage $T-S$ curves, and hence indicates the 
existence of the non-equilibrium and noncritical phase transitions even 
in the isocharge processes. Secondly, the adiabatic 
and isothermal $\Phi-Q$ behaviors are also distinct. 
Lastly, the present model does not allow for the HP transition in the neutral limit. 
There are also some features that are common to the present model and the 
other two universal classes, e.g. the isovoltage $T-S$ curves are all similar, 
and, more importantly, the high temperature limit of the present case 
agree precisely with the cases of the former two classes of theories, 
even up to constant numerical coefficients. 
With further evidences from the study on the cases of Kerr-AdS and Kerr-Newman-AdS 
black holes in Einstein gravity (details not presented here), it appears that 
the recently reported AdS/phonon gas correspondence \cite{hightemp} is universal, 
irrespective of the spacetime dimensions, 
the gravity models, the symmetry of the event horizons as well as the 
amount of charges carried by the black hole solutions. 

\blue{In the course of our series of works on the RPS formalism for black hole thermodynamics, 
we have repeatedly encountered the question on what is meant by a 
variable gravitational coupling constant. The answer is two-folded. On the one hand, 
in a full theory of gravitation in which the quantum features of gravity is considered, 
the gravitational coupling constant can indeed be variable along the renormalization group 
orbit. In such setting, it is natural to consider the effect of variable gravitational 
coupling constant on the macroscopic behavior of black holes as macroscopic objects. 
On the other hand, a variable gravitational coupling constant
is necessary to make the thermodynamic description of black holes extensive, i.e. to
make the Euler homogeneity to hold \cite{wang2021black}. {\em Without Euler homogeneity, 
the thermodynamic properties would become scale dependent. In particular, 
the intensive properties of black holes would become dependent on the size or mass 
of the black hole, which contradicts to the meaning of the
word ``intensive''.} This has long been a problem in black hole thermodynamics, which is
also the underlying reason for us to propose the RPS formalism. }

\blue{In spite of the necessity of considering gravitational coupling constant 
as a variable to make the thermodynamic 
description extensive, there is still a possibility to keep it fixed, such as in 
our observational universe. In such a scenario, the $\mu\rd N$ term in the first law 
could be removed, just like in the thermodynamic description of 
a closed thermodynamic 
system consisted of ordinary matter. Even in such cases, the Euler relation 
\eqref{Eu} must still contain the $\mu N$ term, albeit it becomes a constant term. 
The only consequence of fixing the coupling constant
is to consider the black holes as closed thermodynamic systems. In other word, 
{\em in a universe with fixed gravitational coupling constant, all black holes as 
thermodynamic systems need to be closed.} }

\section*{Acknowledgement}

This work is supported by the National Natural Science Foundation of China under the grant
No. 12275138.

\providecommand{\href}[2]{#2}\begingroup
\footnotesize\itemsep=0pt
\providecommand{\eprint}[2][]{\href{http://arxiv.org/abs/#2}{arXiv:#2}}

\end{document}